\documentclass[aps,prl,twocolumn,preprintnumbers,
superscriptaddress,floatfix]{revtex4}

\usepackage{amssymb,multirow,amsmath}
\usepackage{graphicx,color}

\begin{document}

\newcommand{\Tr}{\mbox{Tr\,}}
\newcommand{\beq}{\begin{equation}}
\newcommand{\eeq}{\end{equation}}
\newcommand{\bea}{\begin{eqnarray}}
\newcommand{\eea}{\end{eqnarray}}
\renewcommand{\Re}{\mbox{Re}\,}
\renewcommand{\Im}{\mbox{Im}\,}

\title{Vacuum Alignment and Phase Structure of Holographic Bi-Layers}

\author{Nick Evans}
\affiliation{STAG Research Centre, School of Physics and Astronomy, University of
Southampton, Southampton, SO17 1BJ, UK }
\email{evans@phys.soton.ac.uk}
\author{Keun-Young Kim} 
\affiliation{School of Physics and Chemistry,
Gwangju Institute of Science and Technology, Gwangju 500-712, Korea}
\email{fortoe@gist.ac.kr}

\begin{abstract}
\noindent  We study the D3/probe D5 system with two domain wall hypermultiplets. The conformal symmetry can be broken by a magnetic field, $B$, (or running coupling) which promotes condensation of the fermions on each individual domain wall. Separation of the domain walls promotes condensation of the fermions between one wall and the other. We study the competition between these two effects showing a first order phase transition when the separation is $\sim 0.56 \lambda^{1/4} B^{-1/2}$. We identify extremal brane configurations which exhibit both condensations simultaneously but they are not the preferred ground state. 
\end{abstract}

\maketitle

\section{Introduction}

Holographic \cite{Maldacena:1997re,Witten:1998qj} brane systems, such as the D3/probe D5 system~\cite{Karch:2000gx,DeWolfe:2001pq,Erdmenger:2002ex} we will study, resemble physical systems such as graphene. Fermionic matter degrees of freedom are isolated on 2+1d surfaces (domain walls) whilst they interact by gauge degrees of freedom that propagate in a 3+1d bulk. Conformal symmetry can be imposed on the brane system by choice of a sufficient amount of supersymmetry, although at the expense of extra scalar and fermion degrees of freedom. Holographic methods apply when the background ${\cal N}=4$ gauge theory is strongly coupled. Graphene is a conformal system of a massless fermion and electromagnetic interactions but whether it is a strongly interacting system remains contentious - the effective electromagnetic constant is larger in the graphene system due to the reduced speed of light in the effective relativistic theory of the fermions on its surface (see for example \cite{Semenoff:2011jf} for a recent discussion of these issues). It may be possible in the future to increase the interaction strength in real materials. The holographic system achieves strong coupling through a choice of large $N$ in a non-abelian gauge theory rather than through a choice of large coupling but the resulting dynamics is most likely comparable. 

The D3/probe D5 system has been studied in detail \cite{Jensen:2010ga,Evans:2010hi} for the case of a single domain wall using the probe approximation \cite{Karch:2002sh,Erdmenger:2007cm}, which we will also employ here (a recent related model can be found in \cite{Filev:2013vka}). The introduction of either temperature or density triggers a first order phase transition to a deconfined fermion plasma phase the moment the conformal symmetry is broken. A more interesting phase diagram results if a magnetic field is applied.  The magnetic field induces condensation of a fermion anti-fermion bilinear ($\langle \bar{f} f \rangle$) generating a mass gap in the system \cite{Filev:2007gb}. Temperature, $T$, and chemical potential, $\mu$, oppose this condensation leading to a critical line in the temperature density plane - the transition is first order with temperature alone, second order in a range of $\mu$ at finite T, and of Berezinskii-Kosterlitz-Thouless (BKT) type at zero temperature with changing $\mu$. The phase diagram can be found in~\cite{Evans:2010hi}.

A different sort of condensation occurs in models with two probe branes separated in the 3+1d space in which the glue live~\cite{Skenderis:2002vf,Karch:2005ms,Davis:2011am,Grignani:2012qz,Chang:2013toa}. Such a configuration is analogous to placing two graphene sheets, each with a massless fermion on its surface, parallel but separated by a short distance (note this is not equivalent to building graphite where the graphene layers are offset and a mass is induced for the surface fermions). Here, in the brane system, at zero $T$ and $\mu$, the separation is the conformal symmetry breaking parameter which triggers condensation. In this case though the condensation is between the fermions on one brane, $f$, with those on the other, $g$ through the operator $\langle \bar{f} g \rangle$. The branes display this symmetry breaking by joining together in the spirit of the Sakai Sugimoto model \cite{Sakai:2004cn}. First experiments of this sort with graphene have been reported in \cite{Gorbachev} and indeed show strong interactions between the layers. 

Here we will be interested in the brane bilayer configuration with a magnetic field (at zero $T$ and $\mu$). The separation and the B field both break the conformal symmetry. They each favour fermion condensation but in different channels. This system is therefore an example of a strongly coupled system with a vacuum alignment problem - which of the two fermion condensates $\langle \bar{f} f \rangle$ and $\langle \bar{f} g \rangle$ will form for different choices of parameters? We explore this system and show that as the separation of the branes grows at fixed $B$ the system undergoes a first order phase transition between vacua characterized by these two condensates. It would have been interesting if a vacuum in which both condensates existed were to form and we do find such brane systems  that are extrema of the effective potential (ie regular brane configurations) but they correspond to a maxima of the effective potential.

\section{The Holographic Dual Theory}

We will loosely represent QED interactions by the large N dynamics of  
${\cal N}=4$ super Yang Mills theory on the surface of a stack of
D3 branes. It is described at zero temperature by AdS$_5\times
S^5$ \cite{Maldacena:1997re,Witten:1998qj}
\beq\label{ads4}
\begin{split}
ds^2  = &  {(\rho^2+L^2) \over R^2} (dx_{2+1}^2 + dz^2)\\
&+ {R^2 \over (\rho^2+L^2)} (d\rho^2 + \rho^2 d \Omega_2^2
+ dL^2 + L^2 d \tilde{\Omega}_2^2) \,,
\end{split}
\eeq
where we have written the geometry to display the
directions the D3 lie in ($x_{2+1},z$). We will embed the D5
on ($x_{2+1}$, $\rho$ and $\Omega_2$) and the transverse directions are $L$
and $\tilde{\Omega}_2$, plus the 3 direction that we call $z$. $R$ is the AdS radius.

We will introduce quenched matter via a probe D5 brane. The matter content is a single
Dirac fermion plus scalar super partners (that will become massive in the presence of any supersymmetry breaking). The 
underlying brane configuration is as follows:
\begin{center}\begin{tabular}{ccccccccccc}
& 0 & 1 & 2 & 3 & 4 & 5 & 6 & 7 & 8 & 9  \\
D3 & - & - & - & - & $\bullet$ & $\bullet$ & $\bullet$ &
$\bullet$ & $\bullet$ & $\bullet$  \\
D5 & - & - & - & $\bullet$ & - & - & - & $\bullet$ & $\bullet$ &
$\bullet$
\end{tabular}\end{center}

The action for the D5, at zero $T$ and $\mu$, is just it's world volume 
\beq \label{action} 
\begin{split}
S &\sim T \int d^6\xi e^\phi \sqrt{- {\rm det} G}  \\ 
  &\sim  \int d\rho~ e^\phi\rho^2 \sqrt{1 +
L^{'2} +   (\rho^2 + L^2)^2 z^{'2}}  \,,
\end{split} 
\eeq 
where $T$ is the tension, $\phi$ the dilaton (which is constant in pure AdS) and we have dropped angular
factors on the two-sphere.  Here we have rescaled each of $L, \rho$ and $z$ by a factor of $R$.

Numerical computation of $-S$ evaluated on a solution gives the vacuum energy. This energy diverges like $\Lambda^3$ for large cut off $\Lambda$. The difference in energy between any two solutions is finite.

\begin{figure}[]
\centering
\includegraphics[width=6.5cm]{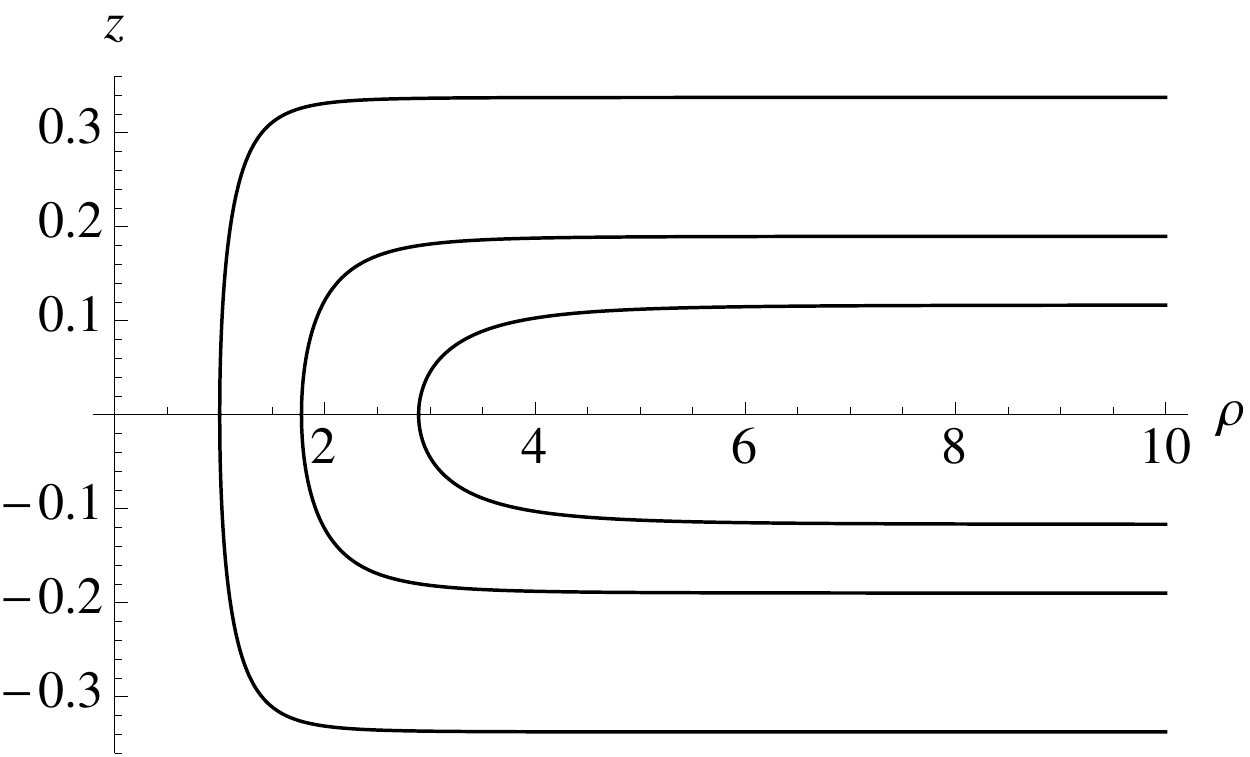} 
\caption{D5 embeddings ($z$ vs $\rho$) with $B=e^\phi=0$ showing $\langle \bar{f} g \rangle$ condensation. Note the larger the 
UV separation of the D5s the deeper the embedding penetrates into AdS as for Wilson loops.  }
\label{justL}
\end{figure}

\noindent {\bf Conformal Mono-Layers:} The embedding that minimizes the action in the case of a single D5 brane in AdS has both $z$ and $L$ constant. $z$ determines the position of the domain wall in the 3+1d bulk. $L$ determines the mass of the quark ($m = L/2 \pi \alpha'$). 

\noindent{\bf Bi-Layer Condensation:}
We now consider a D5 and a $\bar{\mathrm{D}}5$ defect lying parallel but separated by $\Delta z$ in the $z$ 
direction (a similar configuration to that in \cite{Davis:2011am,Grignani:2012qz}).
These represent our two domain walls. 
The separation introduces a conformal symmetry breaking parameter and potentially allows
the strong interactions to generate a fermion condensate between the two branes (we refered to this above as the $\langle \bar{f} g \rangle$ condensate). We seek solutions with $L(\rho)=0$
and a non-trivial $z$ profile in $\rho$. The $z$ embedding equation gives
\begin{equation} 
\partial_\rho \left[{ \rho^6 z'  \over \sqrt{1  + \rho^4 z^{'2}}} \right]  = 0 \,.
\end{equation} 
One can solve this numerically by picking some  $\rho_{0}$ and setting $z'(\rho_{0}) = \infty$ - see Fig \ref{justL} for the solutions. The D5 and $\bar{\mathrm{D}}5$ choose to join at the scale $\rho_0$. Their joining represents the formation of the 
$\langle \bar{f} g \rangle$ condensate which breaks the flavour symmetries of the two branes to the diagonal sub-group. The system experiences a mass gap on the scale $\rho_{0}$.

The analysis and solutions in this case are very similar to the standard computation of Wilson loops in AdS \cite{Maldacena:1998im,Rey:1998bq}. As there, we can use the $z$ independence of the solution to identify a conserved quantity $\Pi_z$ so that
\beq \label{piz} \Pi_z  =  {\rho^6 z' \over \sqrt{1 + \rho^4 z^{'2}}} \,. \eeq
Evaluation at $\rho_0$ (where $z' \rightarrow \infty$), gives $\Pi_z=\rho_0^4$.
The separation of the D5 and $\bar{\mathrm{D}}5$ is given by 
\beq 
\Delta z = {2 \over \rho_0}  \int_1^\infty {dy \over y^2 \sqrt{y^8-1}} 
= \frac{2\sqrt{\pi} \Gamma[5/8]}{\rho_0\Gamma[1/8]} \sim \frac{0.675}{\rho_0} \,.
\eeq
The energy density per unit two volume of the configuration is then given by substituting $z'$ from (\ref{piz}) into the action, giving
\beq 
\begin{split} 
E &= 2 \rho_0^3 \int_1^\infty \left( {y^6 \over \sqrt{y^8-1}}  - y^2 \right)dy \\
& = \rho_0^3 \left(\frac{2}{3}  - \frac{2 \sqrt{\pi} \Gamma[15/8]  \tan[\pi/8] }{7 \Gamma[11/8]}    \right) \sim 0.442 \rho_0^3 \,.
\end{split}
\eeq
Here we have regulated the UV by subtracting the $y^2$ term in the integral as the counter term.
The energy density scales as $1/ (\Delta z)^3$ as expected on dimensional grounds.

\begin{figure}[]
\centering
\includegraphics[width=6.5cm]{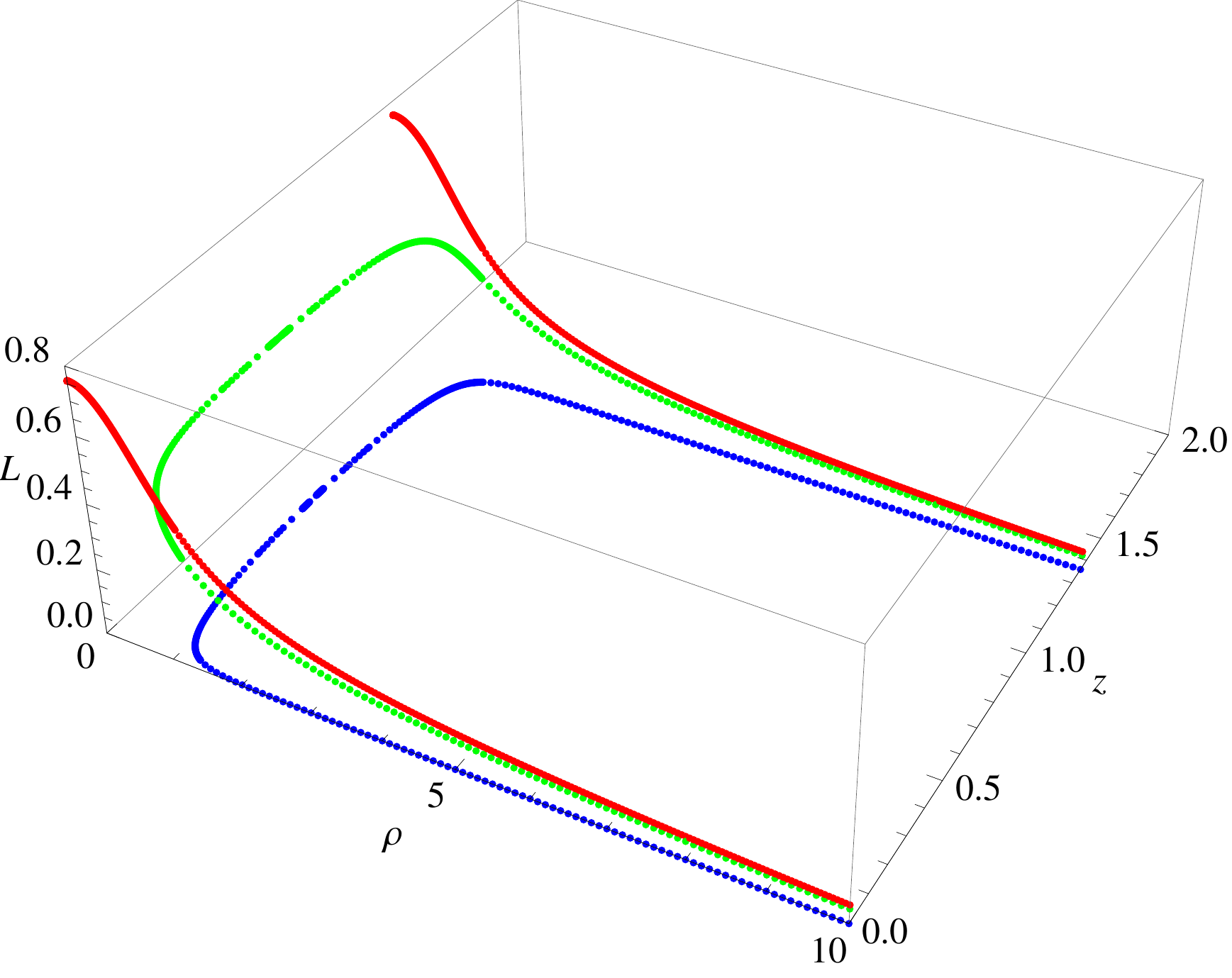} 
\caption{Example D$5/\bar{\mathrm{D}}5$ embeddings (with $B=1$). The coordinates shown are defined in (\ref{ads4})}
\label{embed}
\end{figure}

\noindent {\bf Mono-Layer with a B field:} The DBI action in its full form also contains a gauge field 
\beq S \sim T \int
d^6\xi e^\phi \sqrt{- {\rm det} [G   + 2 \pi \alpha' F]}  \,. \eeq
The gauge field $A^\mu$ in $F$ is dual to operators of the form $\bar{f} \gamma^\mu f$ and their source, a 2+1d U(1)$_B$ baryon number gauge field. We can therefore use $F$ to introduce a fixed background  magnetic field, B \cite{Filev:2007gb,Jensen:2010ga,Evans:2010hi}. 
The resulting action takes the form \eqref{action} if one identifies an effective background dilaton
\beq \label{Bdil} e^\phi = \sqrt{1 + {(2 \pi \alpha' )^2B^2  \over (\rho^2+L^2)^2}} \,. \eeq
Note that the B field is not part of the ${\cal N}=4$ gauge dynamics that are loosely being used to represent the QED interations of graphene. This is, though, a clean method to introduce conformal symmetry breaking on the defect theory. It is interesting that the inclusion of B can be written as an effective non-backreacted dilaton profile (ie an effective running coupling) and one could imagine exploring the dependence of the theory on different choices of that effective dilaton. This approach was explored for 3+1d gauge theories with flavour in~\cite{Evans:2011eu}. Here we will explore the B field case for the defects.

In the far UV (large $\rho$) the embedding Lagrangian is simply ${\cal L}= \rho^2 L'^{2}$ with solutions to the Euler-Lagrange equation of the form \beq \label{asymp}L = m + {c \over \rho} + \cdots \,.
\eeq Here $m$ is a mass term for the fermions  and $c$ the expectation value for $\langle \bar{f} f \rangle$ - note $m$ has dimension one and $c$
dimension two adding to three as required for a Lagrangian term in
2+1d.  Note that when $m=c=0$ the theory is conformal. Including
a non-zero $m$ or $c$ breaks the SO(3) symmetry of $\tilde{\Omega}_2$. Were $c$ to be non-zero when
$m=0$ it would be an order parameter for the spontaneous breaking
of the symmetry.

The solution of the full Euler-Lagrange equation governing the embedding from (\ref{action}) with (\ref{Bdil}) can be found numerically
by shooting out from $\rho=0$ with boundary condition $L'(0)=0$. This picks out the unique asymptotic value of $c$ that gives a regular embedding in the IR.
The solution with $c=0$ is regular in pure AdS$_5$. However in the presence of B there is a another solution that is energetically prefered with non-zero $c$. The red curves in Fig \ref{embed} show this solution which live at fixed $z$. The global symmetry is broken and the fermions have a mass gap of the order of $L(0)$.

\noindent {\bf Bi-Layer with a B field:} We can now turn to the novel, more complex problem of a D5 and a  $\bar{\mathrm{D}}5$ separated in $z$, with a surface magnetic field. Here we expect there to be a vacuum alignment issue between formation of the condensates $\langle \bar{f} f \rangle$ and $\langle \bar{f} g \rangle$.  It is convenient to introduce scaled coordinates $(\rho, L) \rightarrow R \sqrt{2 \pi \alpha' B}  (\rho, L)$ and $z \rightarrow R z/\sqrt{2 \pi \alpha' B}$. The full action is then
\begin{equation} \label{origin}
S \sim \int d \rho ~ \rho^2  \sqrt{1 + {1 \over (\rho^2 + L^2)^2} }   \sqrt{1 +
L^{'2} +   (\rho^2 + L^2)^2 z^{'2}} \,. \eeq
Naively this gives a system of coupled equations for $L(\rho), z(\rho)$ but we use the $z$ independence of the action to find the conserved quantity
\beq \Pi_z = \rho^2  \sqrt{1 + {1 \over (\rho^2 + L^2)^2} } {(\rho^2 + L^2)^2 z^{'} \over  \sqrt{1 +
L^{'2} +   (\rho^2 + L^2)^2 z^{'2}} } \,. \eeq
Note that $\Pi_z$ is again related to properties of the embedding at the turning point where $z' \rightarrow \infty$ but there is no simple interpretation here. We have
\beq \label{z} z^{'2} = { \Pi_z^2 (1 + L^{'2}) \over \rho^4  (\rho^2 +L^2)^2 (1   - \Pi_z^2 + (\rho^2 + L^2)^2)} \,. \eeq
We can now use the Legendre transformed action to find $L(\rho)$ given $\Pi_z$. The Legendre transformed action is
\beq \label{L} S_{LT} \simeq \int d \rho ~ \sqrt{1 + L^{'2}} {\sqrt{\rho^4 ( 1 + (\rho^2 + L^2)^2 )  - \Pi_z^2} \over \rho^2 + L^2} \,. \eeq
There are solutions with $z$ constant which are just two copies of the mono-layer with B field solution at separated $z$ - these are shown in red in Fig \ref{embed}. They represent condensation of the fermions on each brane individually ($\langle \bar{f} f \rangle$ and $\langle \bar{g} g \rangle$). There are also solutions with $L(\rho)=0$
which can be found numerically from (\ref{z}) - each choice of $\Pi_z$ gives a solution with a different separation $\Delta z$. These are similar to the bi-layer condensation discussed above with the D5 and $\bar{\mathrm{D}}5$ joining in the interior of the AdS space. An example is shown in blue in Fig \ref{embed}. The presence of the B field tends to push the junction point to higher $\rho$. They represent $\langle \bar{f} g \rangle$ condensation in the field theory.

\begin{figure}[]
\centering
\hspace{-3mm}
\includegraphics[width=6.7cm]{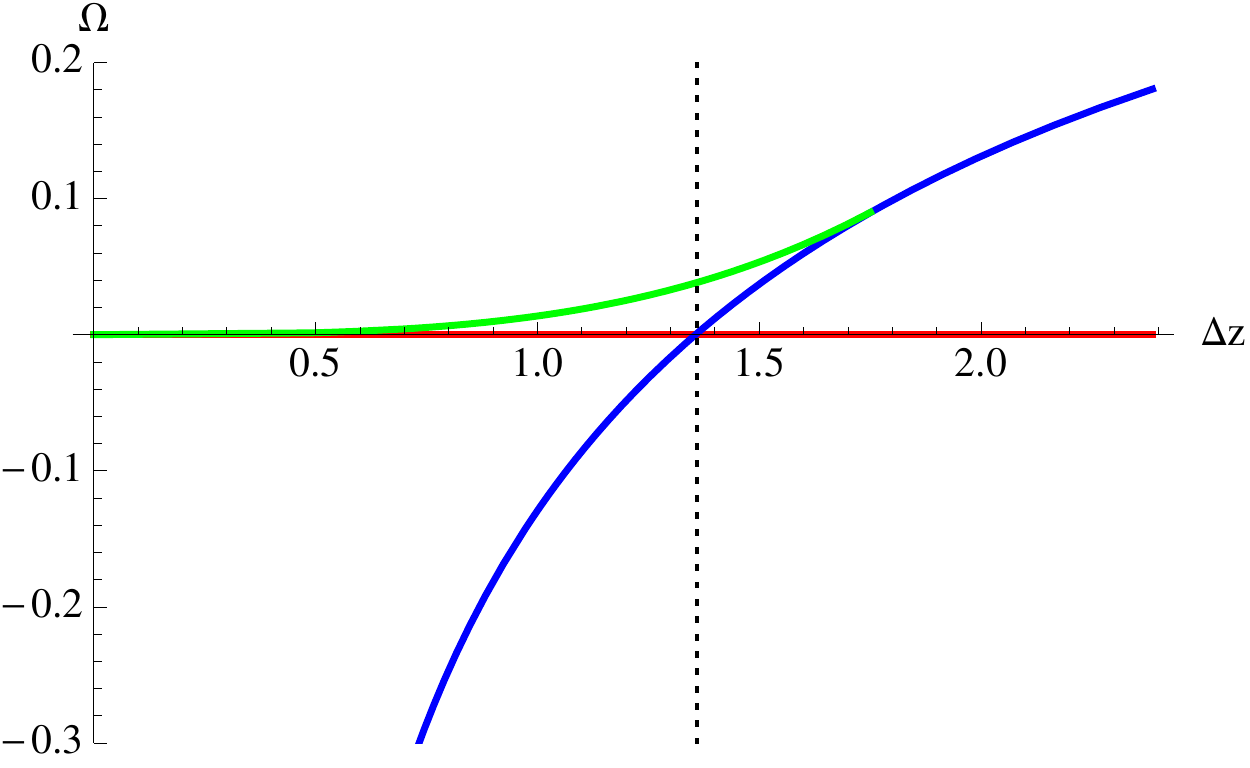} 
\includegraphics[width=6.5cm]{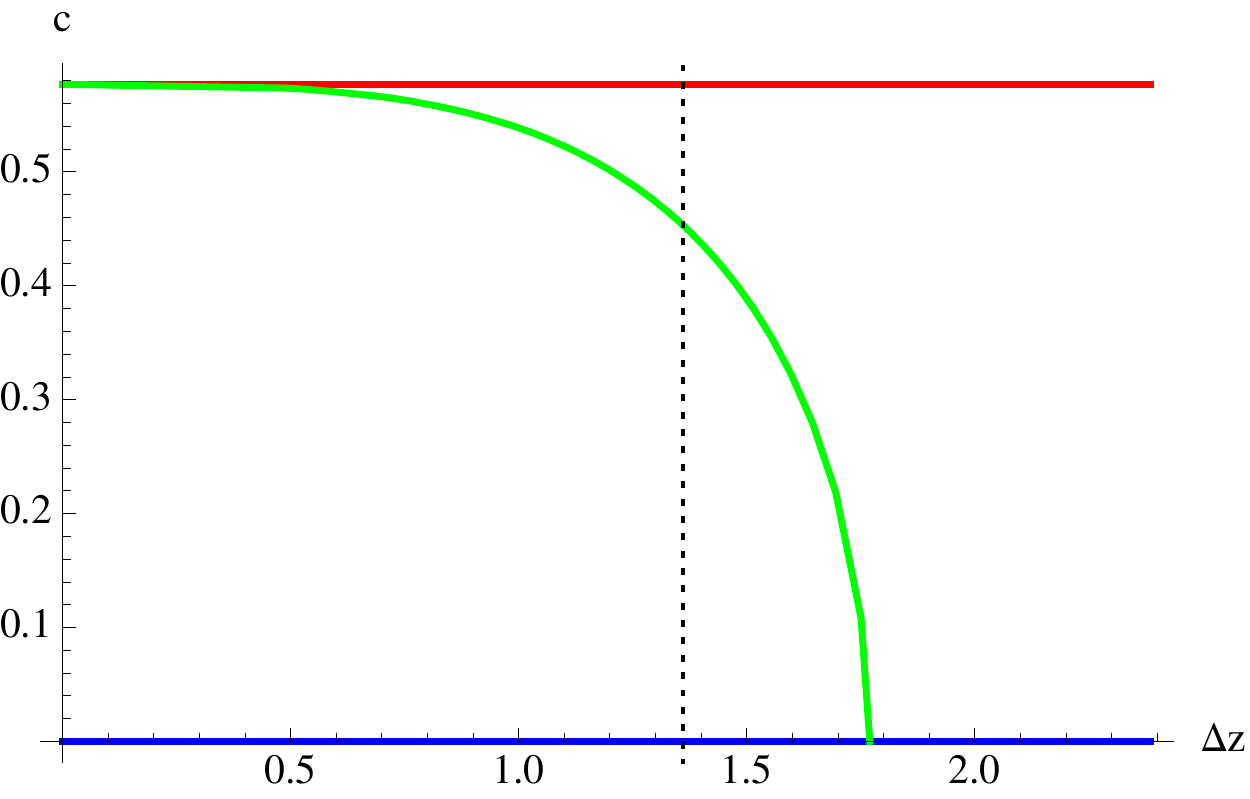} 
\includegraphics[width=6.5cm]{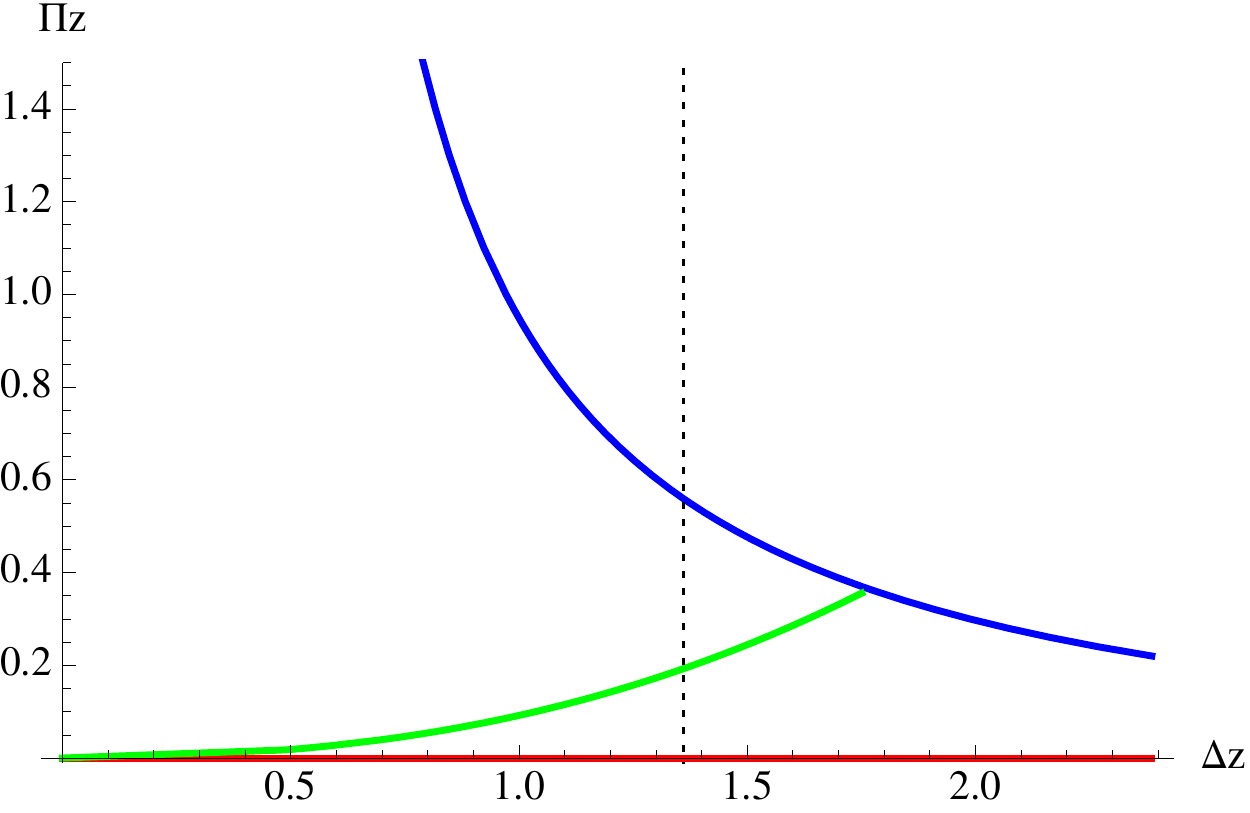}
\caption{Analysis of the solutions of the form shown in Fig \ref{embed} against separation $\Delta z$. Top: The free energy density. Middle: the $\langle \bar{f} f \rangle$ condensate. Bottom: the conserved quantity $\Pi_z$.  }
\label{answer}
\end{figure} 

Finally we can seek solutions with non-trivial $L(\rho)$ and $z(\rho)$. Here, for fixed $\Pi_z$, we find solutions of the equations resulting from (\ref{L}) for $L(\rho)$ subject to the boundary $L'(\rho_0)=0$. We then solve (\ref{z}) for the $z$ embedding in each case. For generic choices of $\rho_0$ the solution for $z(\rho)$ does not satisfy $z'(\rho_0) \rightarrow \infty$ and the solution is not regular. One must therefore scan in $\rho_0$ for the regular embeddings. We show an example of such a regular solution in green in Fig \ref{embed}. Whilst numerically intensive this procedure is straightforward. After collecting such solutions for all choices of $\Pi_z$ one can then compare solutions of all three types (red, green and blue in Fig \ref{embed}) with the same value of $\Delta z$. To compute their vacuum energy we substitute the functions $L(\rho), z(\rho)$ into (\ref{origin}).  Since the energy has a UV divergence we subtract the energy of the disconnected red solution (which is independent of $\Delta z$) to regularize.

The results of our analysis are summarized by the plots in Fig \ref{answer} which are all for the zero mass case ($m=0$ in (\ref{asymp})). The top plot shows the free energy density (relative to the unconnected red embeddings in Fig \ref{embed}) against the separation of the domain walls $\Delta z$. At small separation the linked D5 $\bar{\mathrm{D}}5$ configurations are energetically favourable and the condensate $\langle \bar{f} g \rangle$ forms. At wider separation the disconnected configurations are prefered and the condensation is in the channel $\langle \bar{f} f \rangle$. The transition occurs at $\Delta z \sim 1.4$  - undoing our coordinate rescaling this gives $\Delta z \sim 1.4 R/\sqrt{ 2 \pi \alpha' B} \sim 0.56 \lambda^{1/4}B^{-1/2}.$  In the middle plot we show the $\langle \bar{f} f \rangle$ condensate ($c$ in (\ref{asymp})). Clearly at the transition there is a discrete jump and thus the transition is of first order. In the bottom figure we also show the value of the conserved quantity $\Pi_z$ across the transition. 

The green lines show the values of the free energy, condensate and $\Pi_z$ for the linked solutions with non-zero $z(\rho)$ and $L(\rho)$. These configurations have both $\langle \bar{f} f \rangle$ and $\langle \bar{f} g \rangle$ condensation present. As can be seen from the top graph they are never energetically favoured though. Given we have found all the regular solutions of the system we know the number of turning points of the effective potential and their energetic ordering. So we can deduce the qualitative form of the effective potential for the condensate $c$ for example. We sketch it in Fig \ref{sketch}.  Everything is consistent with the first order phase transition we have identified.

\begin{figure}[]
\centering
\includegraphics[width=6.5cm]{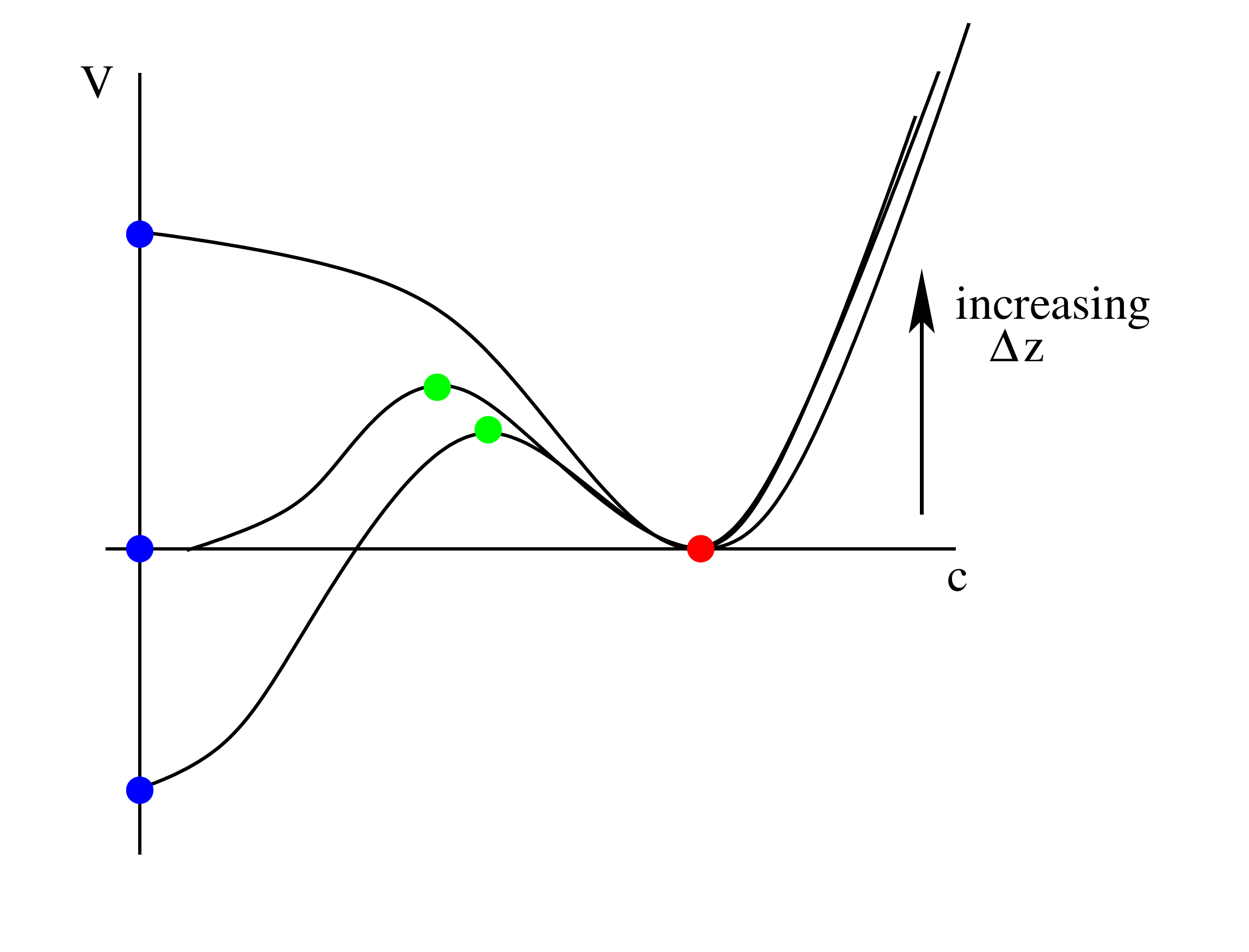} 
\caption{Sketch of the effective potential for the condensate $c$ ($\langle \bar{f} f \rangle$). The coloured points are the extrema corresponding to the regular D5 embedding of the different forms in Fig \ref{embed}.}
\label{sketch}
\end{figure}

\section{Summary}

We have identified a new first order phase transition in holographic bilayer systems. The conformal symmetry of the 
D3/bi-probe-D5 system can be broken by separating the layers or by the presence of a magnetic field. The separation favours condensation of the fermions across the layers. The magnetic field favours condensation of the fermions on each individual layer. We have shown that in the presence of both there is a transition from the former to the latter when the separation grows to $\sim 0.56 \lambda^{1/4} B^{-1/2}$. We also identified regular brane configurations that have both condensates present but they are never the vacuum, instead representing maxima in the effective potential as sketched in Fig \ref{sketch}. Whether this physics can be identified in bi-layer graphene systems or other condensed matter systems in the future remains an interesting and open question. 

\vspace{1cm}

\noindent {\bf Acknowledgements:}

NE is grateful for the support of STFC. KK acknowledges support of an GIST Global University Project. We thank Gordon Semenoff and Namshik Kim for comments on our work.


\begin{thebibliography}{ll}

\bibitem{Maldacena:1997re}
  J.~M.~Maldacena,
  Adv.\ Theor.\ Math.\ Phys.\  {\bf 2} (1998) 231
  [hep-th/9711200].

\bibitem{Witten:1998qj}
  E.~Witten,
  Adv.\ Theor.\ Math.\ Phys.\  {\bf 2} (1998) 253
  [hep-th/9802150].

\bibitem{Karch:2000gx}
  A.~Karch and L.~Randall,
  JHEP {\bf 0106} (2001) 063
  [arXiv:hep-th/0105132].

\bibitem{DeWolfe:2001pq}
  O.~DeWolfe, D.~Z.~Freedman and H.~Ooguri,
  Phys.\ Rev.\  D {\bf 66} (2002) 025009
  [arXiv:hep-th/0111135].


\bibitem{Erdmenger:2002ex}
  J.~Erdmenger, Z.~Guralnik and I.~Kirsch,
  Phys.\ Rev.\  D {\bf 66} (2002) 025020
  [arXiv:hep-th/0203020].

	
\bibitem{Semenoff:2011jf}
  G.~W.~Semenoff,
  Phys.\ Scripta T {\bf 146} (2012) 014016
  [arXiv:1108.2945 [hep-th]].


\bibitem{Jensen:2010ga}
  K.~Jensen, A.~Karch, D.~T.~Son and E.~G.~Thompson,
  Phys.\ Rev.\ Lett.\  {\bf 105} (2010) 041601
  [arXiv:1002.3159 [hep-th]].


\bibitem{Evans:2010hi}
  N.~Evans, A.~Gebauer, K.~-Y.~Kim and M.~Magou,
  Phys.\ Lett.\ B {\bf 698} (2011) 91
  [arXiv:1003.2694 [hep-th]].





	

	
\bibitem{Karch:2002sh}
  A.~Karch and E.~Katz,
  JHEP {\bf 0206} (2002) 043
  [hep-th/0205236].



\bibitem{Erdmenger:2007cm}
  J.~Erdmenger, N.~Evans, I.~Kirsch and E.~Threlfall,
  Eur.\ Phys.\ J.\  A {\bf 35} (2008) 81
  [arXiv:0711.4467 [hep-th]].




\bibitem{Filev:2013vka}
  V.~G.~Filev, M.~Ihl and D.~Zoakos,
  arXiv:1310.1222 [hep-th].



\bibitem{Filev:2007gb}
  V.~G.~Filev, C.~V.~Johnson, R.~C.~Rashkov and K.~S.~Viswanathan,
  JHEP {\bf 0710} (2007) 019
  [hep-th/0701001].
	

	
\bibitem{Skenderis:2002vf}
  K.~Skenderis and M.~Taylor,
  JHEP {\bf 0206} (2002) 025
  [hep-th/0204054].
	
\bibitem{Karch:2005ms}
  A.~Karch, A.~O'Bannon and K.~Skenderis,
  JHEP {\bf 0604} (2006) 015
  [hep-th/0512125].

\bibitem{Davis:2011am} 
  J.~L.~Davis and N.~Kim,
  JHEP {\bf 1206}, 064 (2012)
  [arXiv:1109.4952 [hep-th]].

\bibitem{Grignani:2012qz}
  G.~Grignani, N.~Kim and G.~W.~Semenoff,
  Phys.\ Lett.\ B {\bf 722} (2013) 360
  [arXiv:1208.0867 [hep-th]].
	
\bibitem{Chang:2013toa}
  H.~-C.~Chang,
  arXiv:1310.5734 [hep-th].

\bibitem{Sakai:2004cn}
  T.~Sakai and S.~Sugimoto,
  Prog.\ Theor.\ Phys.\  {\bf 113} (2005) 843
  [hep-th/0412141].
	
	\bibitem{Gorbachev}
  R. V. Gorbachev et al.,
  Nature Phys {\bf 8}, (2012) 896
  [arXiv:1206.6626].
	
\bibitem{Maldacena:1998im}
  J.~M.~Maldacena,
  Phys.\ Rev.\ Lett.\  {\bf 80} (1998) 4859
  [hep-th/9803002].
	
\bibitem{Rey:1998bq}
  S.~-J.~Rey, S.~Theisen and J.~-T.~Yee,
  Nucl.\ Phys.\ B {\bf 527} (1998) 171
  [hep-th/9803135].
	



\bibitem{Evans:2011eu}
  N.~Evans, A.~Gebauer, M.~Magou and K.~-Y.~Kim,
  J.\ Phys.\ G {\bf 39} (2012) 054005
  [arXiv:1109.2633 [hep-th]].

\end{thebibliography}
\end{document}